\documentclass[]{aa}
\usepackage{graphicx}
\usepackage{txfonts}

\begin{document}

%%%%%%%%%%%%%%%%%%%%%%%%%%%%%%%%%%%%%%%%%%%%%%%%

\title{A solar twin in the eclipsing binary LL~Aqr.}

\titlerunning{The eclipsing binary LL~Aqr}
\author{D.~Graczyk\inst{1,2,3},
R.~Smolec\inst{3},
K.~Pavlovski\inst{4},
J.~Southworth\inst{5},
G.~Pietrzy\'nski\inst{3,2},
P. F. L.~Maxted\inst{5},
P.~Konorski\inst{6},\\
W.~Gieren\inst{2,1},
B.~Pilecki\inst{3},
M.~Taormina\inst{3},
K.~Suchomska\inst{6},
P.~Karczmarek\inst{6},
M.~G{\'o}rski\inst{1,2},
\and
P.~Wielg{\'o}rski\inst{3}
}
\authorrunning{D. Graczyk et al.}
\institute{Millenium Institute of Astrophysics, Santiago, Chile
\and Universidad de Concepci\'on, Departamento de Astronom\'ia, Casilla 160-C, Concepci\'on, Chile
\and Nicolaus Copernicus Astronomical Center, Polish Academy of Sciences, ul. Bartycka 18, 00-716 Warszawa, Poland
\and Department of Physics, University of Zagreb, Bijeni{\'c}ka cesta 32, 10000 Zagreb, Croatia
\and Astrophysics Group, Keele University, Keele, Staffordshire, ST5 5BG, UK
\and Warsaw University Observatory, Al. Ujazdowskie 4, 00-478 Warsaw, Poland
}

\abstract
{}
{In the course of a project to study eclipsing binary stars in vinicity of the Sun, we found that the cooler component of LL~Aqr is a solar twin candidate. This is the first known star with properties of a solar twin existing in a non-interacting eclipsing binary, offering an excellent opportunity to fully characterise its physical properties with very high precision.}
{We used extensive multi-band, archival photometry and the Super-WASP project and high-resolution spectroscopy obtained from the HARPS and CORALIE spectrographs. The spectra of both components were decomposed and a detailed LTE abundance analysis was performed. The light and radial velocity curves were simultanously analysed with the
Wilson-Devinney code. The resulting highly precise stellar parameters were used for a detailed comparison with PARSEC, MESA, and {\sc garstec} stellar evolution models.}
{LL~Aqr consists of two main-sequence stars (F9\,V + G3\,V) with masses of $M_1 = 1.1949\pm0.0007$ and $M_2=1.0337\pm0.0007$~M$_\odot$, radii $R_1 = 1.321\pm0.006$ and $R_2=1.002\pm0.005$~R$_\odot$, temperatures $T_1=6080\pm45$ and $T_2=5703\pm50$ K and solar chemical composition [M/H] $=0.02\pm 0.05$. The absolute dimensions, radiative and photometric properties, and atmospheric abundances of the secondary are all fully consistent with being a solar twin. Both stars are cooler by about 3.5$\sigma$ or less metal abundant by 5$\sigma$ than predicted by standard sets of stellar evolution models. When advanced modelling was performed, we found that full agreement with observations can only be obtained  for values of the mixing length and envelope overshooting parameters that are hard to accept. The most reasonable and physically justified model fits found with MESA and {\sc garstec} codes still have discrepancies with observations but only at the level of 1$\sigma$. The system is significantly younger that the Sun, with an age between 2.3 Gyr and 2.7 Gyr, which agrees well with the relatively high lithium abundance of the secondary, $A({\rm Li}) = 1.65\pm0.10$~dex.

}
{}


\maketitle

\section{Introduction}

Solar twins are of special astrophysical interest. As the stars most physically similar to the Sun, they are prime targets for extrasolar planet search projects \citep[e.g.][]{but98,how10,bed15}, allow for precise differential abundance analysis relative to the Sun \citep[e.g.][]{mel09,ram09,gon10}, enable the determination of the colours of the Sun \citep[e.g.][]{cas12,ram12}, and aid the calibration of the effective temperature scale \citep{cas10}, to mention a few important areas.

Detailed spectroscopic analysis of bright solar twins can give precise temperatures, gravities, and surface abundances, however, their radii and especially masses can be found with much less precision. The closest solar twin, 18~Sco, had its physical radius directly determined using optical interferometry \citep[][precision of 1\%]{baz11}. However, its mass was measured only indirectly using asteroseismology and the homology relation, giving a precision of 3\% \citep{baz11}. Using similar methodology the solar twins of the 16~Cyg binary system had their radii and masses determined with precisions of about 2\% and 4\%, respectively \citep{whi13}. HIP 56948, the star most similar to the Sun identified to date, has had its mass and radius determined to a precision of 2\% but only indirectly, i.e.\ utilizing stellar evolution models and making differential isochrone analysis \citep{mel12}. In a number of recent more general studies of solar twins \citep[e.g.][]{por14,ram14,nis15} the determination of their masses (and subsequently radii) was performed only by means of comparison with theoretical evolutionary tracks.

In this work we report a detailed analysis of LL~Aqr (HD~213896, HIP~111454, $\alpha_{2000}\!=\!22^h 34^m 42^s\!\!.2$, $\delta_{2000}\!=\!-03^{\circ} 35^{'} 58^{''}$), a well detached eclipsing binary containing two solar-type stars. We included this system in our long-term project of investigating nearby eclipsing binaries with {\it Hipparcos} parallaxes and/or a large angular separation between the components \citep{gra15,gall16}. LL~Aqr was previously analysed by \cite{iba08}, who obtained the first photometric solution and absolute dimensions from dedicated $UBV$ photometry and spectroscopy; later by \cite{gri13}, who obtained a refined orbital solution of the system; and subsequently by \cite{sou13}, who carried out a comprehensive study of the light and velocity curves. While checking the consistency of published radiative parameters of this system, we found that the stars are substantially cooler than the literature estimates and the secondary probably falls into the solar-twin region. This was a prime motivation for our reanalysis of LL~Aqr. Here, for the first time, the fully model-independent derivation of the radius and mass of a solar twin is presented. Both the physical parameters are derived with very high precision of much better than 1\%, giving us an excellent opportunity to investigate the universality of commonly used stellar evolution isochrones in studies of solar twins and solar-type stars.

The outline of the paper is as follows: section~\ref{observ} gives some details of data used, section~\ref{analys} describes the method of analysis, section~\ref{twin} contains a comparison of the secondary in LL~Aqr with the Sun, section~\ref{evol} is devoted to a detailed comparison of LL~Aqr with evolutionary models, and the section~\ref{fin} presents some concluding remarks.

 \section{Observations}\label{observ}

\subsection{Photometry}\label{photo}
\subsubsection{UBV}

We used broadband Johnson $UBV$ photometry collected by \cite{iba08} using two telescopes at Ege University Observatory in Turkey. The data comprise 1925 photometric points in each band. The details of the observations and set-up are given in \cite{iba08}.

\subsubsection{WASP}

We used an extensive light curve of LL~Aqr obtained by the Super-WASP consortium in the course of their search for transiting extrasolar planets \citep{pol06}. The light curve was cleaned of outliers using 3$\sigma$ clipping and details are given in \cite{sou13}. Because the cleaned light curve contains 21\,362 datapoints and it is much larger then the combined Johnson photometry, we decided to reduce the number of datapoints as follows. We cut out datapoints covering both eclipses, in the orbital phase intervals $-0.01$ to $0.01$ (412 points) and $0.305$ to $0.330$ (668 points), using ephemeris given by \cite{sou13}. The remaining datapoints were used to calculate the mean out-of-eclipse magnitude. Then we selected every 20th datapoint from these remaining datapoints in such a way that their average was closest to the out-of-eclipse magnitude. Finally we combined the data from the outside and inside eclipses and obtained a final WASP light curve with 2104 datapoints. This procedure introduces no bias in the results because the light curve of LL~Aqr is an almost perfectly flat outside eclipse.

\subsection{Spectroscopy}

\subsubsection{HARPS}

We obtained spectra of LL~Aqr  with the High Accuracy Radial velocity Planet Searcher \citep[HARPS;][]{may03} on the European Southern Observatory 3.6 m telescope in La Silla, Chile. LL~Aqr is a bright target so observations were generally obtained in marginal observing conditions or during twilight. Observations were obtained between 2008 December 10 and 2014 September 8. A total of 16 spectra were secured in high efficiency (``EGGS'') mode. The exposure times were typically 260\,s resulting in an average signal to noise per pixel S/N $\sim$ $55$. All spectra were reduced on-site using the HARPS Data Reduction Software (DRS).

\subsubsection{CORALIE}

Fifteen spectra were obtained with the CORALIE spectrograph on the Swiss 1.2 m Euler Telescope, also at La Silla observatory, between 2008 October 7 and 2009 September 16. The exposure times were about 600\,s giving a typical S/N near 5500\AA\ of 40 per pixel. The spectra were reduced on-site using the automated data reduction pipeline.

\section{Analysis}
\label{analys}
\subsection{Radial velocities \label{rad_vel}}

\begin{table}
\centering
\caption{Radial velocity measurements for LL~Aqr. Index ``1'' denotes the hotter (primary) star and ``2'' denotes the cooler
(secondary)  component. Numbers in brackets give the uncertainty. BJD means Barycentric Julian Date.}
\label{tab_rv}
\begin{tabular}{@{}lrrrr@{}}
\hline \hline
BJD & $RV_1$ &  $RV_2$  & Spectr. \\
-2450000& (km s$^{-1}$) &(km s$^{-1}$) & \\
\hline
4747.61886 & 21.800(38) & $-$46.260(78) & CORALIE \\
4810.51861 & 25.655(35) & $-$50.728(72) & HARPS \\
4819.52699 & $-$69.371(38) & 59.147(78) & CORALIE \\
4820.53294 & $-$54.560(38) & 42.030(80) & CORALIE \\
4821.52471 & $-$38.150(38) & 23.003(79) & CORALIE \\
4822.52530 & $-$23.327(30) & ---        & CORALIE \\
5060.81786 & $-$74.108(35) & 64.757(72) & HARPS \\
5086.61574 & $-$2.934(34) & $-$17.511(76) & CORALIE \\
5086.78175 & $-$1.319(33) & $-$19.235(68) & CORALIE \\
5087.57081 &  5.248(40) & $-$26.947(83) & CORALIE \\
5087.72071 &  6.394(40) & $-$28.260(84) & CORALIE \\
5088.56206 & 12.123(40) & $-$34.901(83) & CORALIE \\
5088.68656 & 12.863(39) & $-$35.773(81) & CORALIE \\
5089.56985 & 17.581(39) & $-$41.204(82) & CORALIE \\
5089.67527 & 18.020(39) & $-$41.784(83) & CORALIE \\
5090.56685 & 21.594(40) & $-$45.921(81) & CORALIE \\
5090.66626 & 21.945(39) & $-$46.282(81) & CORALIE \\
5120.58442 & $-$67.357(35) & 56.988(71) & HARPS \\
5144.51605 & $-$35.991(27) & 20.699(52) & HARPS \\
5447.67953 & $-$28.600(26) & 12.189(51) & HARPS \\
5470.52337 &  1.656(35) & $-$22.847(72) & HARPS \\
5471.50987 &  9.192(35) & $-$31.583(71) & HARPS \\
5479.64355 & 12.602(34) & $-$35.511(70) & HARPS \\
5503.50550 & $-$58.374(35) & 46.575(71) & HARPS \\
5504.49921 & $-$73.214(34) & 63.691(71) & HARPS \\
5504.62207 & $-$73.806(34) & 64.401(71) & HARPS \\
5535.51853 & 24.345(34) & $-$49.085(71) & HARPS \\
6211.50396 & $-$72.532(34) & 62.766(71) & HARPS \\
6241.51587 & 23.754(34) & $-$48.531(69) & HARPS \\
6242.51240 & 25.395(34) & $-$50.442(70) & HARPS \\
6908.66171 & 25.659(35) & $-$50.572(71) & HARPS \\
\hline
\end{tabular}
\end{table}

\begin{table}
\begin{center}
\caption{Observed magnitudes of the LL~Aqr system}
\label{magnitud}
\begin{tabular}{ccccl}
\hline \hline
\multicolumn{4}{c}{Band} & Ref.  \\
 Original  &  & Transformed & &  \\
\hline
U &9.850(62)&&& 1\\
B$_{\rm T}$ &9.978(28)&B&9.872(32)& 3\\
B &9.765(30)&&& 1\\
B &9.831(37)&&& 4\\
B &9.917(156)&&& 5\\
B &9.822(107)&&& 6\\
V$_{\rm T}$ &9.386(23)&V&9.330(29)& 3 \\
V &9.206(26)&&& 1\\
V &9.230(33)&&& 4\\
V &9.229(20)&&& 5\\
V &9.252(76)&&& 6\\
J$_{\rm 2MASS}$ &8.145(23)& J$_{\rm J}$&8.193(24)&2\\
H$_{\rm 2MASS}$ &7.872(33)& H$_{\rm J}$&7.872(36)&2\\
K$_{\rm 2MASS}$ &7.819(23)& K$_{\rm J}$&7.840(24)&2\\
\hline
\end{tabular}
\\
\end{center}
{\small Source: 1 --- \cite{iba08}, 2 --- \cite{cut03}, 3 --- Tycho-2 \citep{hog00}, 4 --- {\it Hipparcos} (ESA 1997), 5 --- AAVSO \citep{hen16}, 6 --- URAT1 \citep{zac15}\\
}
\end{table}

We used RaveSpan \citep{pil12,pil13} to measure the radial velocities of both stars in the system using the broadening function (BF) formalism \citep{ruc92,ruc99}. We used templates from the library of synthetic LTE spectra by \cite{col05} matching the mean values of estimated effective temperature and gravity of the stars in the binary. The abundance was assumed to be solar. There is a small difference in the systemic velocities of the two components, where the primary component is blueshifted by $195 \pm 12$\,m\,s$^{-1}$ with respect to the secondary. A very similar difference of $270 \pm 140$\,m\,s$^{-1}$ was reported by \cite{sou13} based on velocimetry from \cite{gri13}. However our absolute systemic velocity of the system is different by as much as 1.4 km s$^{-1}$, which is a value that is much larger than the measurements errors. It is very unlikely that this offset of 1.4 km\,s$^{-1}$ is a true shift because we do not see any progressive trend in the systemic velocity of the system. Most probably this shift comes from different methods of radial velocity determination used in our work and by \cite{gri13}. The systemic velocity reported by \cite{iba08} is, on the other hand, fully consistent with our value.

The line profiles of both stars are virtually Gaussian, suggesting the rotational velocities are small. The total broadening of the lines interpreted in terms of projected equatorial rotational velocity is $6.5\pm1.0$\,km\,s$^{-1}$ and $5.7\pm0.8$\,km\,s$^{-1}$, for the primary and secondary components, respectively. In reality the projected rotational velocities are smaller than this because the total line broadening includes contributions from macroturbulence, microturbulence, and instrumental broadening. Decomposition of those effects were carried out during atmospheric analysis of disentangled spectra (see Sec.~\ref{abu}). The primary is about 2.5 times more luminous in V band than the secondary, and although they have similar rotational broadening, the root-mean-square (rms) of the radial velocity residuals is similar for both components. This is in contrast to expectations because one would obtain higher S/N ratio and more precise radial velocity measurements for a brighter star. Some additional variability in the primary, for example the presence of small amplitude non-radial pulsations may cause larger than expected rms. 

\subsection{Spectral decomposition\label{deco}}

\begin{figure}
\includegraphics[angle=0,scale=.50]{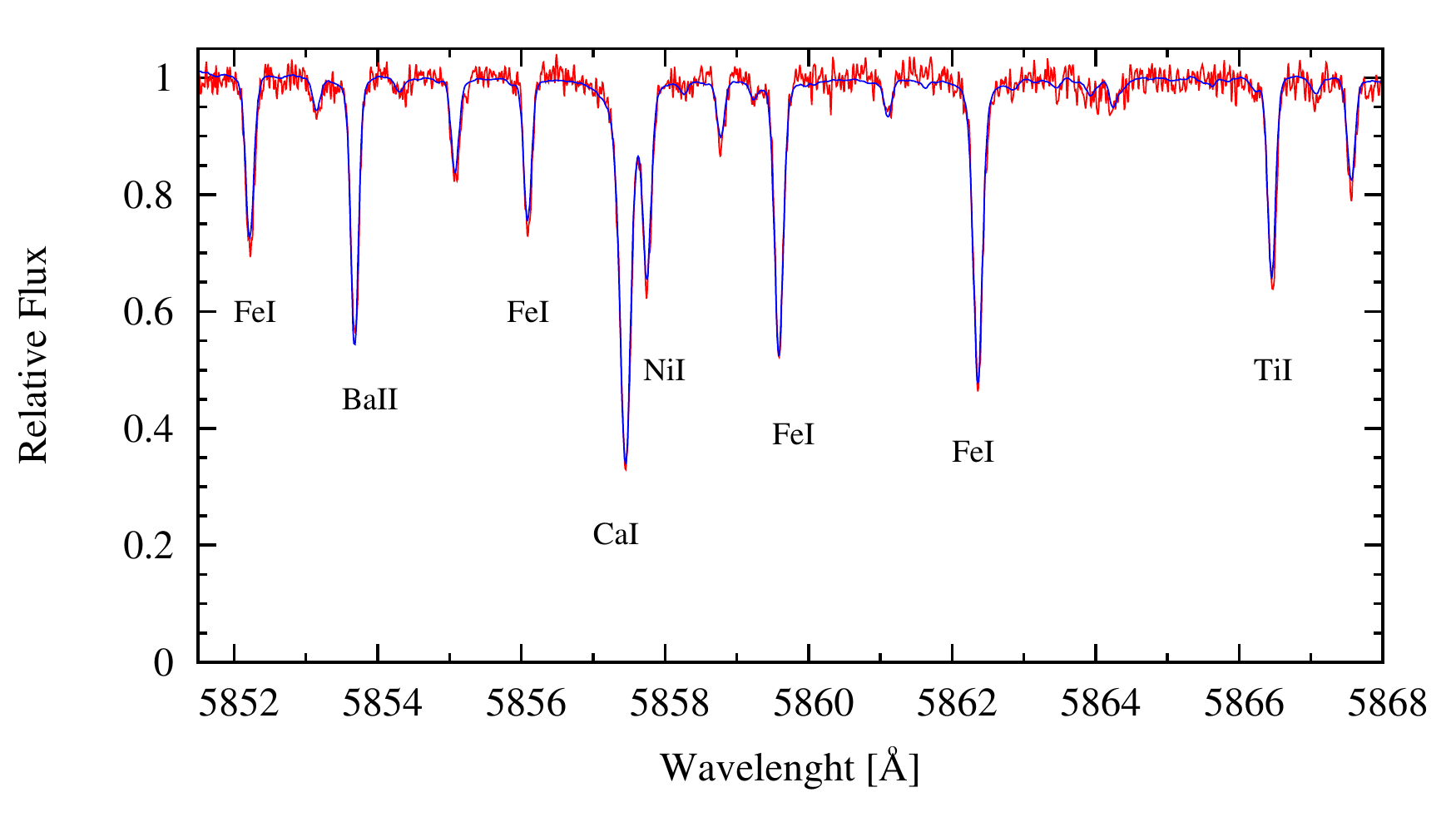}
\caption{Comparison of the decomposed secondary spectrum (red line, S/N $=$ 90) with the solar spectrum (blue line, S/N $=$ 700)
taken by UVES with a similar resolving power. Some absorption lines are labelled. The spectra are very similar. \label{fig:spec}}
\end{figure}

During the secondary (shallower) eclipse, the light from the cooler component is completely blocked by the companion. A spectrum taken exactly during the mid-point of the secondary eclipse would contain only light from the primary. As none of our high-resolution spectra were taken at this orbital phase, we decided to decompose the observed spectra into individual spectra of both components. We included all HARPS and CORALIE spectra in this analysis and used the iterative method outlined by \cite{gon06}, which is implemented in the RaveSpan code. We used the previously measured radial velocities and we repeated the iterations twice. In order to renormalise the spectra we used the photometric parameters from \cite{sou13} but with temperatures lower by about 500\,K (see Sect.~\ref{temp}), and we calculated appropriate light ratios with the Wilson-Devinney code (hereafter WD; see Sec.~\ref{wd} for details and references). The resulting individual spectra have S/N $=$ 165 for the primary and S/N $=$ 90 for the secondary at 5500\,\AA. A comparison of the secondary's spectrum with that of the Sun\footnote{\texttt{http://www.eso.org/observing/dfo/quality/UVES/\\pipeline/solar\_spectrum.html}} is shown in Fig.~\ref{fig:spec}, which intentionally covers the same wavelength interval as the middle panel of Fig.~1 in \cite{mel12}. The spectra are very similar; the secondary's spectrum has some absorption lines that are a little deeper, reflecting its slightly lower surface temperature than the Sun.

\subsection{Atmospheric and abundance analysis}\label{abu}

\begin{table*}
\centering
\caption{\label{tab:abund} Measured photospheric elemental abundances for the components of LL~Aqr, derived from
our disentangled HARPS and CORALIE spectra. Columns list the atomic number, element, and degree of ionisation,
and then for each component the logarithmic value of the elemental abundance on the usual scale in which
$\log n(H) = 12$, the number of spectral lines measured, and the logarithmic abundance relative to the Sun of
element X with respect to hydrogen. The last column gives the reference photospheric solar values from \cite{a09}.}
\begin{tabular}{rlccccccccc} \hline
A & Species        &     Star\,A     & Lines & [X/H]$_{\rm A}$ &  Star\,B        & Lines& [X/H]$_{\rm B}$ & $\log \epsilon_\odot$ \\
\hline
3  & \ion{Li}{i}    & $2.88 \pm 0.10$ &   1 &                   & $1.65 \pm 0.10$ &   1 &                  & $1.05 \pm 0.10$  \\
6  & \ion{C}{i}     & $8.55 \pm 0.07$ &   4 & $ 0.12 \pm 0.09$  & $8.57 \pm 0.06$ &   9 & $0.14  \pm 0.08$ & $8.43 \pm 0.05$ \\
11 & \ion{Na}{i}    & $6.27 \pm 0.07$ &   4 & $ 0.03 \pm 0.08$  & $6.28 \pm 0.05$ &   8 & $0.04  \pm 0.07$ & $6.24 \pm 0.04$  \\
12 & \ion{Mg}{i}    & $7.56 \pm 0.07$ &   2 & $-0.01 \pm 0.08$  & $7.56 \pm 0.06$ &   5 & $-0.04 \pm 0.07$ & $7.60 \pm 0.04$ \\
14 & \ion{Si}{i}    & $7.55 \pm 0.05$ &  13 & $ 0.04 \pm 0.06$  & $7.60 \pm 0.03$ &  14 & $0.09  \pm 0.04$ & $7.51 \pm 0.03$ \\
20 & \ion{Ca}{i}    & $6.38 \pm 0.05$ &  18 & $ 0.04 \pm 0.06$  & $6.38 \pm 0.06$ &  21 & $0.04  \pm 0.07$ & $6.34 \pm 0.04$ \\
21 & \ion{Sc}{ii}   & $3.20 \pm 0.04$ &   5 & $ 0.05 \pm 0.05$  & $3.16 \pm 0.03$ &   9 & $0.01  \pm 0.05$ & $3.15 \pm 0.04$ \\
22 & \ion{Ti}{i}    & $4.93 \pm 0.04$ &  33 & $-0.02 \pm 0.06$  & $4.97 \pm 0.06$ &  37 & $0.02  \pm 0.08$ & $4.95 \pm 0.05$ \\
23 & \ion{V}{i}     & $3.96 \pm 0.05$ &  15 & $ 0.03 \pm 0.09$  & $3.96 \pm 0.06$ &  13 & $0.03  \pm 0.10$ & $3.93 \pm 0.08$ \\
24 & \ion{Cr}{i}    & $5.65 \pm 0.04$ &  11 & $ 0.01 \pm 0.05$  & $5.62 \pm 0.08$ &  29 & $-0.02 \pm 0.09$ & $5.64 \pm 0.08$ \\
25 & \ion{Mn}{i}    & $5.43 \pm 0.04$ &   7 & $ 0.00 \pm 0.06$  & $5.46 \pm 0.08$ &   5 & $0.03  \pm 0.10$ & $5.43 \pm 0.05$ \\
26 & \ion{Fe}{i}    & $7.54 \pm 0.03$ & 229 & $ 0.04 \pm 0.05$  & $7.55 \pm 0.07$ & 229 & $0.05  \pm 0.08$ & $7.50 \pm 0.04$ \\
26 & \ion{Fe}{ii}   & $7.53 \pm 0.03$ &  16 & $ 0.03 \pm 0.05$  & $7.54 \pm 0.06$ &  16 & $0.04  \pm 0.08$ & $7.50 \pm 0.04$ \\
27 & \ion{Co}{i}    & $4.90 \pm 0.05$ &   6 & $-0.09 \pm 0.08$  & $4.99 \pm 0.07$ &  22 & $0.00  \pm 0.10$ & $4.99 \pm 0.07$ \\
28 & \ion{Ni}{i}    & $6.25 \pm 0.04$ &  81 & $ 0.03 \pm 0.06$  & $6.27 \pm 0.08$ &  88 & $0.05  \pm 0.09$ & $6.22 \pm 0.04$ \\
29 & \ion{Cu}{}     & $4.24 \pm 0.12$ &   2 & $ 0.05 \pm 0.13$  & $4.22 \pm 0.10$ &   4 & $0.03  \pm 0.11$ & $4.19 \pm 0.04$  \\
39 & \ion{Y}{ii}    & $2.22 \pm 0.08$ &   6 & $ 0.01 \pm 0.09$  & $2.18 \pm 0.03$ &  10 & $-0.03 \pm 0.06$ & $2.21 \pm 0.05$ \\
56 & \ion{Ba}{ii}   & $2.02 \pm 0.07$ &   2 & $ 0.16 \pm 0.11$  & $2.29 \pm 0.07$ &   1 & $0.11  \pm 0.12$ & $2.18 \pm 0.09$ \\
\hline \end{tabular} \end{table*}

A detailed spectroscopic analysis of the disentangled spectra of the components makes possible an independent determination of the effective temperatures for both stars and, in turn, the measurements of the elemental photospheric abundances and metallicity.

The iron lines, and in particular neutral iron lines, \ion{Fe}{i}, are the most numerous in the spectra of both components, and they alone can be used to determine the atmospheric parameters. The condition of excitation balance was used to measure the effective temperature, $T_{\rm eff}$. The microturbulent velocity, $v_t$, was determined by enforcing no dependence between the iron abundance and the reduced equivalent widths, $\log ({\rm EW}/\lambda)$. Here, we used the advantage that very precise surface gravities are available from the radii and masses measured from the light and velocity curves. The values we used are $\log g_{\rm A} = 4.274\pm0.004$, and $\log g_{\rm B} = 4.451\pm0.004$.

Equivalent widths of \ion{Fe}{i} and \ion{Fe}{ii} lines carefully selected from the line list of Bruntt et al.\ (2012) were measured with the {\sc uclsyn} code (Smalley et al.\ 2001), which was also used for the calculation of the theoretical spectra. We derived
$v\sin{i}$ by an optimal fitting of selected lines with rotationally broadened theoretical spectra. We account for an instrumental profile of the HARPS spectrograph by measuring the width of telluric lines in original spectra. 
The $T_{\rm eff}$ and $v_t$ were iteratively modified until there were no trends of \ion{Fe}{i} abundance with excitation potential or equivalent width. The uncertainties in $T_{\rm eff}$, and $v_t$ were calculated from the uncertainties in functional dependences of the iron abundance on excitation potential and reduced equivalent widths, respectively. Using the excitation balance method, we obtained $T_{\rm eff,A} = 6080\pm45$ K, $T_{\rm eff,B} = 5690\pm60$ K, $v_{t,A} = 1.20\pm0.08$\,km\,s$^{-1}$\,,  and $v_{t,B} = 1.15\pm0.11$\,km\,s$^{-1}$. The iron abundances from the most numerous \ion{Fe}{i} lines are [Fe/H]$_{\rm A} = 0.04\pm0.05$ and [Fe/H]$_{\rm B} = 0.05\pm0.08$. In the calculation of the uncertainties in abundances we take into account the error propagation due to uncertainties in the $T_{\rm eff}$ and $v_t$ besides an intrinsic scatter in the abundances for different lines. We therefore find that the iron abundances for both components of LL~Aqr are indistinguishable from solar.

Determination of the iron abundance from singly ionised iron lines, \ion{Fe}{ii}, serves as a check for the fulfilment of the iron ionisation balance. The iron abundances derived for both ions and for both stars in the LL~Aqr system are given in Table~\ref{tab:abund}. The iron ionisation balance is fulfilled for both stars, and deviations of iron abundance from two ions are only 0.01\,dex for both stars, which is well within the uncertainties.

The low projected rotational velocities of both stars makes possible high-precision equivalent width measurements and abundance determinations for an additional 16 species besides iron and including lithium (see Sect.~\ref{twin}). The results are given in Table~\ref{tab:abund}. For the average metal abundance relative to solar, and excluding Li and Ba which are based on a single line measurement, we find [M/H]$_{\rm A} = 0.02\pm0.04$ and [M/H]$_{\rm B} = 0.03\pm0.06$. These corroborate our conclusions from the iron abundances that the photospheric compositions of the stars in the LL~Aqr are basically solar to within their 1$\sigma$ uncertainties.

\subsection{Interstellar extinction\label{red}}

We used extinction maps \citep{sch98} with recalibration by \cite{sch11} to determine the reddening in the direction of LL~Aqr. The total foreground reddening in this direction is $E(B-V) = 0.044 \pm 0.003$\,mag. We followed the procedure described in detail in \cite{such15}, assuming a distance to LL~Aqr of $D=0.14$ kpc \citep{sou13} and obtaining $E(B-V)_{LL~Aqr}=0.018\pm 0.020$\,mag, where we assume a very conservative uncertainty.

We also used a calibration between the equivalent width of the interstellar absorption sodium D1 line and reddening by \cite{mun97}. The interstellar D1 line has one narrow component of constant radial velocity of $-4.6$\,km\,s$^{-1}$ with a mean equivalent width of 0.079\,\AA. This corresponds to a colour excess of $E(B-V)=0.018$\,mag and we assumed an error of 0.02\,mag. Both estimates point towards a small reddening in the direction of LL~Aqr. We adopted an extinction of $E(B-V)=0.018 \pm 0.014$\,mag as the final value.

The above reddening is much smaller than the value derived by \cite{iba08} and used in subsequent analysis of the system by \cite{sou13}. The likely reason for the strong overestimate of extinction by \cite{iba08} is that they used the Q method, which for solar-type stars gives large uncertainties because of heavy line blanketing in the wavelength region covered by the $U$, $B,$ and $V$ bands \citep[e.g.][]{wil62}. The ``line-blanketing'' vectors largely coincide with the direction of the reddening vector on the $B\!-\!V$, $U\!-\!B$ diagram, which produces a strong degeneracy between metallicity and reddening, for example\ compare Fig.~3 and Fig.~5 from \cite{arp61} and \cite{wil62}, respectively. Thus even relatively small photometric errors translate into significant reddening uncertainty.

\subsection{Determination of effective temperatures \label{temp}}

The revised value of the extinction causes large change in the dereddened colours of the system, which become significantly redder and suggest much lower temperatures than those derived by \cite{iba08}. Our initial estimate suggested temperatures cooler by about 500\,K, thus placing the secondary in the solar twin region and prompting our reanalysis of this system. We used four different methods to determine the temperature of the stars utilizing (1) colour-temperature calibrations, (2) a calibration of line depth ratio versus temperature, (3) atmospheric analysis of decomposed spectra, and (4) the temperature ratio determined through analysis of the multi-band light curves with the WD code (see Section~\ref{wd} for details).

\subsubsection{Colour -- temperature calibrations}

To estimate the effective temperatures of the two stars, we collected multi-band apparent magnitudes of the system. They are summarised in Table~\ref{magnitud}. Using published Johnson $V$ and $B$ magnitudes, we calculated weighted means that were employed in the temperature determination $V=9.243\pm0.037$\,mag and $B=9.821\pm0.052$\,mag. Both magnitudes have relatively large errors because of the significant spread of the published values. We converted 2MASS magnitudes into Johnson magnitudes using transformation equations from \cite{bes88} and \cite{car01}\footnote{\texttt{http://www.astro.caltech.edu/$\sim$jmc/2mass/v3/\\transformations/}}. The reddening (Sect.~\ref{red}) and the mean Galactic interstellar extinction curve from \cite{fit07}, assuming $R_V=3.1$, were combined with light ratios from the WD code to determine the intrinsic colours of the components. We used a number of colour -- temperature calibrations for a few colours, i.e. $B\!-\!V$ \citep{alo96,flo96,ram05,gon09,cas10}, $V\!-\!J$, $V\!-\!H$ \citep{ram05,gon09,cas10}, and $V\!-\!K$ \citep{alo96,hou00,ram05,mas06,gon09,cas10,wor11}. As the source for infrared photometry is 2MASS \citep{cut03} (see Table~\ref{magnitud}) we used appropriate colour transformations for each calibration. The resulting temperatures were averaged for each colour used and are reported in Table~\ref{tab:temp}.

\begin{table}
\begin{center}
\caption{Temperature of components}
\label{tab:temp}
\begin{tabular}{cll}
\hline \hline
 & \multicolumn{2}{c}{Temperature [K]} \\
Method & Primary & Secondary  \\
\hline
{Colour-temperature calibration} & &\\
$B\!-\!V$ & $6139\pm200$  & $5733\pm177$ \\
$V\!-\!J$ & --& $5747\pm109$\\
$V\!-\!H$ & --& $5692\pm96$\\
$V\!-\!K$ & $6133\pm96$ & $5759\pm87$\\
&&\\
{Line depth ratios} & $6035\pm50$&$5705\pm81$ \\
{Atmospheric analysis} &$6080\pm45$ &$5690\pm60$ \\
&&\\
Weighted mean of above& 6070 & 5713 \\
{Adopted}       & 6080$^1$ & 5703$^2$ \\
\hline
\end{tabular}
\end{center}
$^1$ From the atmospheric analysis\\
$^2$ From the light curve analysis\\
\end{table}

\subsubsection{Line depth ratios}

The method based on ratios of absorption lines with different excitation potentials was claimed to give very precise relative temperatures and thus is well suited to follow temperature changes over the surface of the stars \citep[e.g.][]{gray94}. We used calibrations given by \cite{kov03} that are valid for F-K dwarfs. We measured line depths by fitting Gaussians to unblended line profiles in the decomposed spectra. This way we could use 43 line depth ratios for the primary and 68 line depth ratios for the secondary from the total number of 105 calibrations provided by \cite{kov03}. The derived temperatures are reported in Table~\ref{tab:temp}.

\subsubsection{Adopted values}

The results of the atmospheric analysis are provided in Sect.~\ref{abu}. Table~\ref{tab:temp} summarises our temperature determinations and we can conclude that different methods give very consistent results. In case of the primary its most robust temperature comes from the atmospheric analysis and we fixed this value in the subsequent light curve analysis. The temperature of the secondary was adopted from the light curve analysis that gives a precise value of the temperature ratio $T_2/T_1$ (see Section~\ref{wd}) and it is very close to a weighted mean of 5713\,K. Although temperature -- colour relations suggest slightly higher temperatures for both components, they lie well within the uncertainties of the adopted temperatures. The temperatures we derive are significantly lower than those reported by \cite{iba08}. Their estimate was based on temperature -- colour relations and, as they used too strong an extinction value (see our Sect.~\ref{red}), the dereddened colours were too blue and the temperatures were overestimated.

\subsection{Analysis of the combined light and radial velocity curves \label{wd}}

The motivation for this work was to derive very precise physical parameters of LL~Aqr. This would be possible by combining the orbital parameters from our radial velocities with the precise photometric parameters derived by \cite{sou13} with the {\sc jktebop} code \citep{sou04a,sou04b}. Although this approach is entirely acceptable, we decided to obtain a complete simultaneous solution of the photometry and velocities with another code: the Wilson-Devinney program \citep{wil71} (hereafter WD). This  would allow us to take full advantage of the simultaneous solution of the multi-band light and radial velocity curves and to access additional information about possible systematics in the model. The analysis was performed with version 2007 of the WD code \citep{wil79,wil90,van07}\footnote{\texttt{ftp://ftp.astro.ufl.edu/pub/wilson/lcdc2007/}} equipped with a {\sc python} wrapper written by P.~Konorski. The systematic error caused by simultaneous solution with the WD code seems to be negligible as was extensively discussed in our previous work on the eclipsing binary HD 187669 \citep{hel15}.

\subsubsection{Initial parameters}

We fixed the temperature of the primary component during analysis to $T_1 = 6080$\,K (see Section~\ref{temp}) and the metallicity to [Fe/H] $=0$. The grid size was set to $N=40$ and standard albedo and gravity brightening coefficients for convective stellar atmospheres were chosen. We assumed a detached configuration in the model and a simple reflection treatment (MREF $=$ 1 and NREF $=$ 1). The stellar atmosphere option was used (IFAT $=$ 1), tidal corrections were automatically applied to radial velocity curves, and no flux-level-dependent weighting was used. We assumed synchronous rotation for both components. The epoch of the primary minimum was set according to the ephemeris given by \cite{sou13}. Both the logarithmic limb-darkening law (Klinglesmith \& Sobieski 1970), with coefficients tabulated by \cite{VHa93}, and the linear law, with adjusted linear coefficients, were used during analysis. The starting point for the parameters of the binary system was based on the results in \cite{sou13}.

\subsubsection{Fitting model parameters}

\begin{figure*}
\hspace*{-13pt}
\includegraphics[angle=0,scale=0.52]{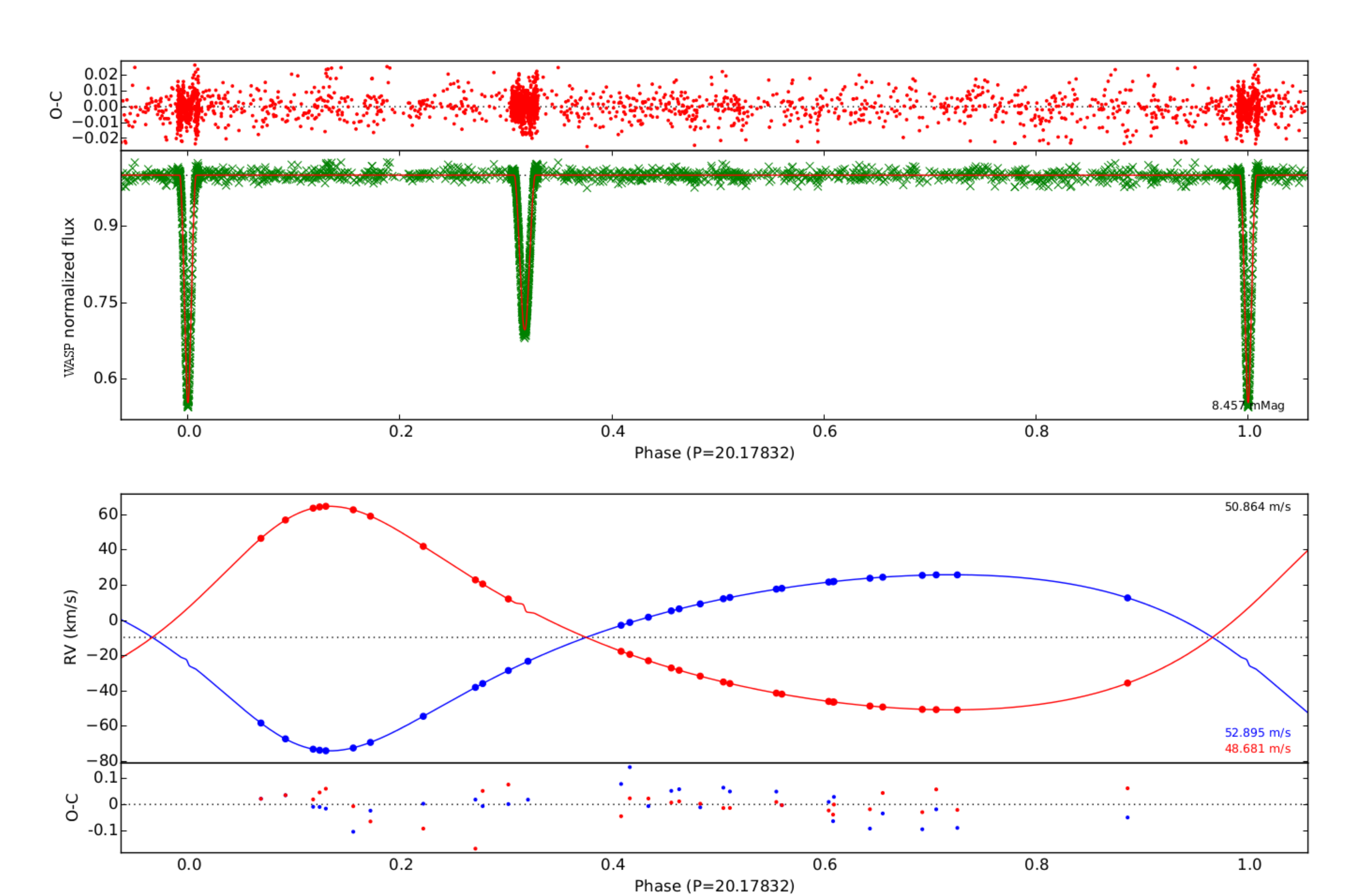}
\caption{Wilson-Devinney model fit to WASP observations (upper panel) and radial velocities (lower
panel). The numbers in the lower right corners are the $rms$ of the fit. \label{fig:sol}}
\end{figure*}

With the WD model we fitted four light curves ($U,B,V$ and WASP) and two radial velocity curves corresponding to both components simultaneously. Each observable curve was weighted only by its $rms$ through comparison with a calculated model curve. Altogether we adjusted the following parameters during our analysis: the orbital period $P_{\rm orb}$, semimajor axis $a$, mass ratio $q$, both systemic radial velocities $\gamma_{1,2}$, phase shift $\phi$, eccentricity $e$, argument of periastron $\omega$, orbital inclination $i$, temperature of the secondary $T_2$, modified Roche potentials $\Omega_{1,2}$, corresponding to fractional radii $r_{1,2}$, and luminosity parameter $L_1$. Additionally we fitted linear law limb darkening coefficients $x$ in four passbands for each component and the third light $l_3$. Our test solutions with the third light as a free parameter invariably returned values that were negative or consistent with zero, so we decided to fix $l_3=0$ during final analysis. The fit to the modified WASP light curve and radial velocities is shown in Fig.~\ref{fig:sol}.

\begin{table}
\begin{centering}
\caption{Results of the final WD analysis for LL~Aqr.}
\label{tab_par_orb}
\begin{tabular}{lcccc}
\hline \hline
Parameter & Primary & Secondary \\
\hline
\multicolumn{3}{l}{Orbital parameters} \\
$P_{\rm orb}$ (d) & \multicolumn{2}{c}{20.178321(3)}  \\
T$_0$ (d) & \multicolumn{2}{c}{2455100.56106(79)}  \\
T$_{\rm p}$ (d) & \multicolumn{2}{c}{2455098.54955(14)}  \\
T$_{\rm s}$ (d) & \multicolumn{2}{c}{2455104.95939(31)}  \\
$a \sin{i}\, (R_\odot)$ & \multicolumn{2}{c}{$40.743(7)$}   \\
$q=M_2/M_1$ & \multicolumn{2}{c}{0.8651(3)}    \\
$\gamma$ (km~s$^{-1}$) & $-$9.81(1) & $-$9.61(1)  \\
$e$ & \multicolumn{2}{c}{0.31654(7)} \\
$\omega$ (deg) & \multicolumn{2}{c}{155.50(4)}  \\
&&&\\
\multicolumn{3}{l}{Photometric parameters} \\
$ i$ (deg) & \multicolumn{2}{c}{89.548(26)}  \\
$T_2/T_1$ & \multicolumn{2}{c}{0.9380(36)}\\
$\Omega$ &32.11(14)&36.70(17)\\
$r_{\rm mean}$ &0.03243(15)&0.02460(12)\\
$T_{\rm eff}$ (K) &6080$^{a}$&5703(21) \\
$L_2/L_1$(U)&\multicolumn{2}{c}{0.3301(9)}\\
$L_2/L_1$(B) &\multicolumn{2}{c}{0.3830(7)}\\
$L_2/L_1$(V) &\multicolumn{2}{c}{0.4174(7)}\\
$L_2/L_1$(WASP) &\multicolumn{2}{c}{0.4358(7)}\\
$x$(U) &0.743(29)&0.869(74) \\
$x$(B) &0.723(22)&0.786(53)\\
$x$(V) &0.538(25)&0.642(42)\\
$x$(WASP) &0.654(24)&0.660(41)\\
&&&\\
\multicolumn{3}{l}{Derived quantities} \\
$a (R_\odot)$ & \multicolumn{2}{c}{$40.744(7)$}   \\
$K$ (km~s$^{-1}$) & 49.948(13) & 57.736(14) \\
RV $rms$ (m~s$^{-1}$) & 53 & 49  \\
U $rms$ (mmag) &\multicolumn{2}{c}{29}\\
B $rms$ (mmag) &\multicolumn{2}{c}{13}\\
V $rms$ (mmag) &\multicolumn{2}{c}{12} \\
WASP $rms$ (mmag) &\multicolumn{2}{c}{8.5}\\
\hline
\end{tabular}
\end{centering}
\\$^a$ Fixed value from atmospheric analysis.
\end{table}

\begin{table}
\centering
\caption{Physical parameters of the components of LL~Aqr}
\label{par_fi}
\begin{tabular}{lccc}
\hline \hline
Parameter& Primary& Secondary  \\
\hline
Spectral type$\,^a$ & F9 V &  G3 V \\
$M$ ($M_{\sun}$) & 1.1949(7)& 1.0337(7) \\
$R$ ($R_{\sun}$) & 1.321(6)& 1.002(5)\\
$\log{g}$ (cgs) & 4.274(4) & 4.451(4) \\
$T_{\rm eff}$ (K) & 6080(45)& 5703(50)\\
$L$ ($L_{\sun}$) & 2.15(7)& 0.958(35) \\
$\upsilon \sin{i}$ (km s$^{-1}$)& 3.5(5)  &  3.6(4) \\
$\upsilon_{\rm t}$ (km s$^{-1}$) & 1.20(8) &1.15(11) \\
$\upsilon_{\rm macro}$ (km s$^{-1})\,^b$ & 4.7(3)& 3.2(2) \\
$[{\rm M}/{\rm H}]$ (dex) &0.02(4) & 0.03(6) \\\hline
MESA Age (Gyr)& \multicolumn{2}{c}{$2.29$ to $2.67$}\\
\textsc{garstec} Age (Gyr)& \multicolumn{2}{c}{$3.7$}\\
Photometric distance (pc) & \multicolumn{2}{c}{134(4)} \\
$E(B\!-\!V)$ (mag) & \multicolumn{2}{c}{0.018(14)}\\
\hline
\end{tabular}
\\$^a$ From $T_{\rm eff}$ using the Pecaut \& Mamajek (2013) calibration.
\\$^b$ From the Gray (2005) calibration for main-sequence stars.
\end{table}

\subsection{Results and physical parameters}

In Table~\ref{tab_par_orb} we present parameters of the final fit, where $T_{0}$, $T_{\rm p}$, $T_{\rm s}$ are epochs of a periastron passage, the primary minimum and secondary minimum, respectively, and $L_2/L_1$ denotes light ratio. The photometric parameters from our simultanous solution are very similar to those reported by \cite{sou13} from modelling the WASP light curve only. They are also consistent within the errors found in his final solution, as expected because we used the same photometric datasets. The model residuals of light curves in both eclipses are in practice almost indistinguishable from the residuals presented in Figures 2 and 3 in \cite{sou13} and we do not repeat them here. The orientation and shape of the orbit ($i$, $e$, $\omega$) are also perfectly consistent, but there is a difference in the velocity semiamplitudes $K_{1,2}$ which are $2\sigma$ larger in our solution. The difference comes from different radial velocity datasets; \cite{sou13} used velocimetry published by \cite{gri13} whilst we used our own extensive and high-precision velocity measurements. We tried to incorporate velocimetry of \cite{gri13} into our solution but it gave residuals that were larger than our velocimetry 
by a factor of 10 and degraded the fit.

The absolute dimensions of the system were calculated using astrophysical constants following \cite{tor10} and additionally a total solar irradiance of $1360.8\pm0.5$ W m$^{-2}$ \citep{kop11}. The resulting effective temperature of the Sun is $T_\odot=5772 \pm 1$ K, which we assumed for calculation of the bolometric luminosities. The parameters of the system are summarised in Table~\ref{par_fi}.

\begin{figure}
\includegraphics[angle=0,scale=.50]{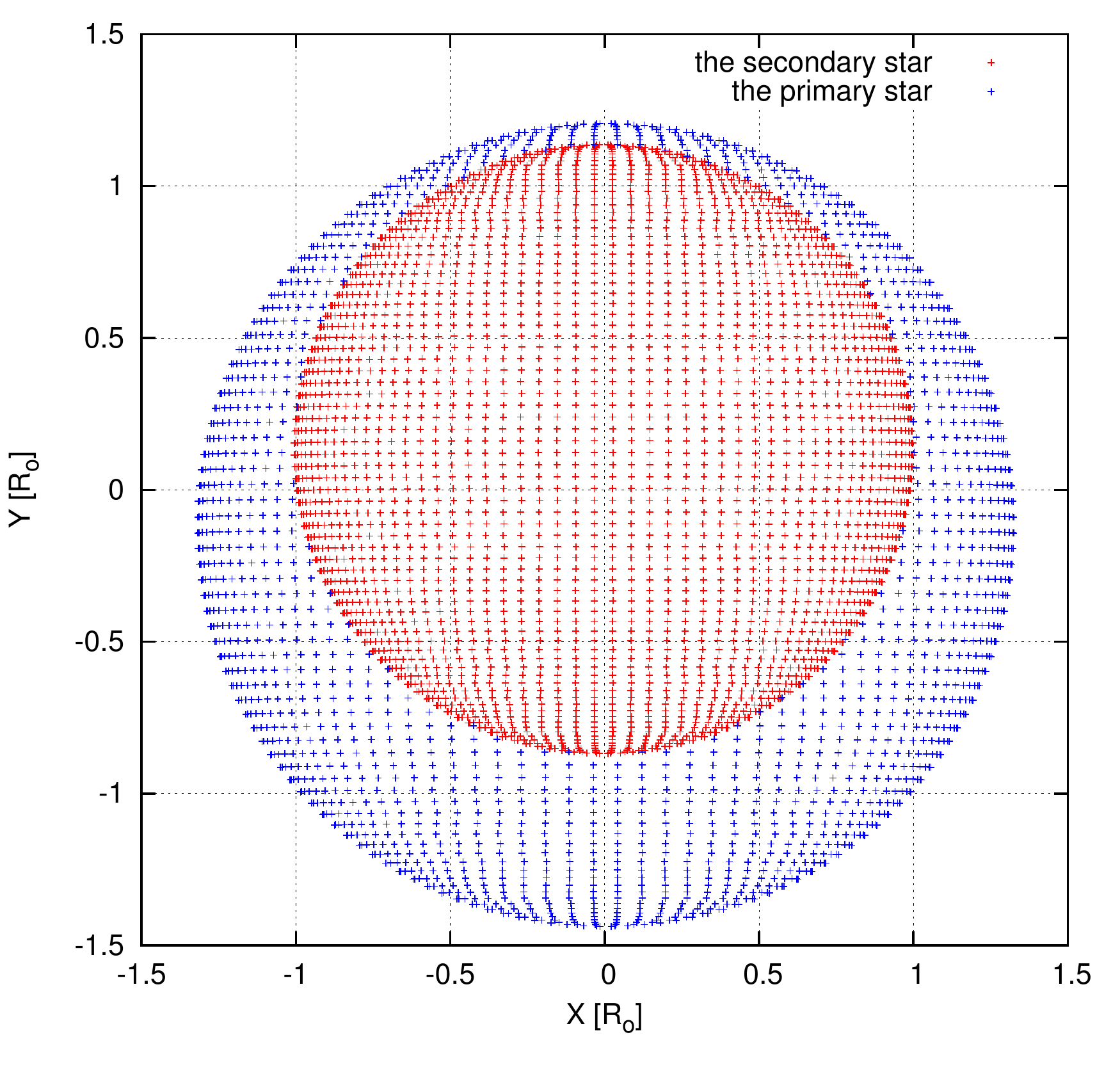}
\caption{Configuration of the system projected onto the sky at the mid-point of the primary eclipse.
The cooler G3 component is transiting the disc of the larger and hotter F9 companion star. The grid is
expressed in solar radii. The actual distance between stars is 32.4 R$_\sun$. \label{fig:primary}}
\end{figure}

The primary eclipse is annular, i.e.\ the cooler component transits the disc of the larger and hotter companion star (see Fig.~\ref{fig:primary}),. The secondary eclipse is a partial eclipse but is almost total, as at mid-eclipse; the primary component covers 99.94\% of the projected surface area of the companion and blocks 99.97\% of the flux in the $V$ band.

\begin{table}
\begin{center}
\caption{Intrinsic colours of the LL~Aqr system}
\label{tab:col}
\begin{tabular}{lrllc}
\hline \hline
Colour & Primary & Secondary & Sun & Ref. \\
\hline
$(U\!-\!B)$ & $-$0.023(80)& 0.128(80)& 0.159(9)& 1\\
$(B\!-\!V)$ &0.528(53) &0.640(54)&0.653(3)&  1\\
$(V\!-\!J)_{\rm 2MASS}$ & 1.001(52)&1.173(52)&1.198(5)& 2\\
$(V\!-\!H)_{\rm 2MASS}$ & 1.247(61)&1.484(61)&1.484(9)&2\\
$(V\!-\!K)_{\rm 2MASS}$ &1.296(55) &1.535(55)&1.560(8)&2\\
$(J\!-\!H)_{\rm 2MASS}$&0.246(40)&0.311(41)&0.286&2\\
$(J\!-\!K)_{\rm 2MASS}$&0.295(34)&0.362(34)&0.362&2\\
$(H\!-\!K)_{\rm 2MASS}$&0.049(40)&0.051(41)&0.076&2\\
\hline
\end{tabular}
\\
\end{center}
{\small Reference to the Sun's colours: 1 -- \cite{ram12}, 2 -- \cite{cas12}}
\end{table}

\section{The secondary as a solar twin} \label{twin}

It is interesting to compare the secondary star of LL~Aqr with the Sun. The differences (LL~Aqr~B $-$ Sun) amount to $-68$\,K in $T_{\rm eff}$, 0.013\,dex in $\log{g}$, 0.02\,dex in $[{\rm M}/{\rm H}],$ and 0.16\,km\,s$^{-1}$ in $\upsilon_{\rm t}$., The two stars are extremely similar, so LL~Aqr~B is one of the best candidates for a solar twin and is certainly the one with the best-known absolute dimensions. The spectra of both stars are very similar although the $6707.8~{\rm \AA}$ lithium line of LL~Aqr~B is stronger than in the Sun, signifying a younger age \citep[see e.g.][]{gal16}. Using Eq.~1 from \cite{car16} and the lithium abundance of $A({\rm Li})=1.65\pm0.10$ dex (LTE), we estimate the age of LL~Aqr~B to be $2.0\pm0.1$ Gyr, which compares well with the isochronal age of between 2.29 and 2.67 Gyr derived from stellar evolution modelling (see Sect.~\ref{evol}). We also compared the intrinsic colours of both components with the Sun (see Table~\ref{tab:col}). Within the quoted errors all colours of LL~Aqr~B are fully consistent with the Sun's colours. The activity of both stars in the LL~Aqr system is very low, in fact undetectable \citep{sou13}, which suggests that the secondary is a quiet star, and so is also in this respect similar to the Sun. Thus all discrimination methods used to establish close kindred with the Sun (i.e.\ similarity of spectra, atmospheric parameters, intrinsic colours, and absolute dimensions) give a very consistent picture for LL~Aqr~B as a solar twin.

\section{Stellar evolution models for LL~Aqr}\label{evol}

With precisely determined stellar parameters, LL~Aqr is an excellent testbed for stellar evolution theory. Its modelling should be relatively simple: it is a well detached system composed of two low-mass main-sequence stars. The secondary is a solar twin, its interior is radiative, and a significant convective zone is present in the envelope. The mass of the primary, on the other hand, puts it in the transition region in which the convective core appears and the convective envelope shrinks \citep[see e.g.\ Fig.~22.7 in][]{kww.book}. We first confronted the parameters of LL~Aqr with the PAdova and TRieste Stellar Evolution Code (PARSEC) isochrones \citep{parsec}, then with stellar evolutionary tracks computed with  the Modules for Experiments in Stellar Astrophysics (MESA) \citep[e.g.][]{mesa3}, and last with the standard set of assumptions (e.g.\ no diffusion, fixed mixing-length parameter calibrated on the Sun). Then, we explored the effects of advanced and non-standard effects, i.e.\ of element diffusion and different convective efficiencies (mixing-length parameters) for the two components using the MESA stellar evolution code and, finally, we performed a Bayesian analysis using \textsc{garstec} code.

\subsection{PARSEC isochrones and MESA standard evolutionary models}

\begin{figure}
\centering
\includegraphics[width=9cm]{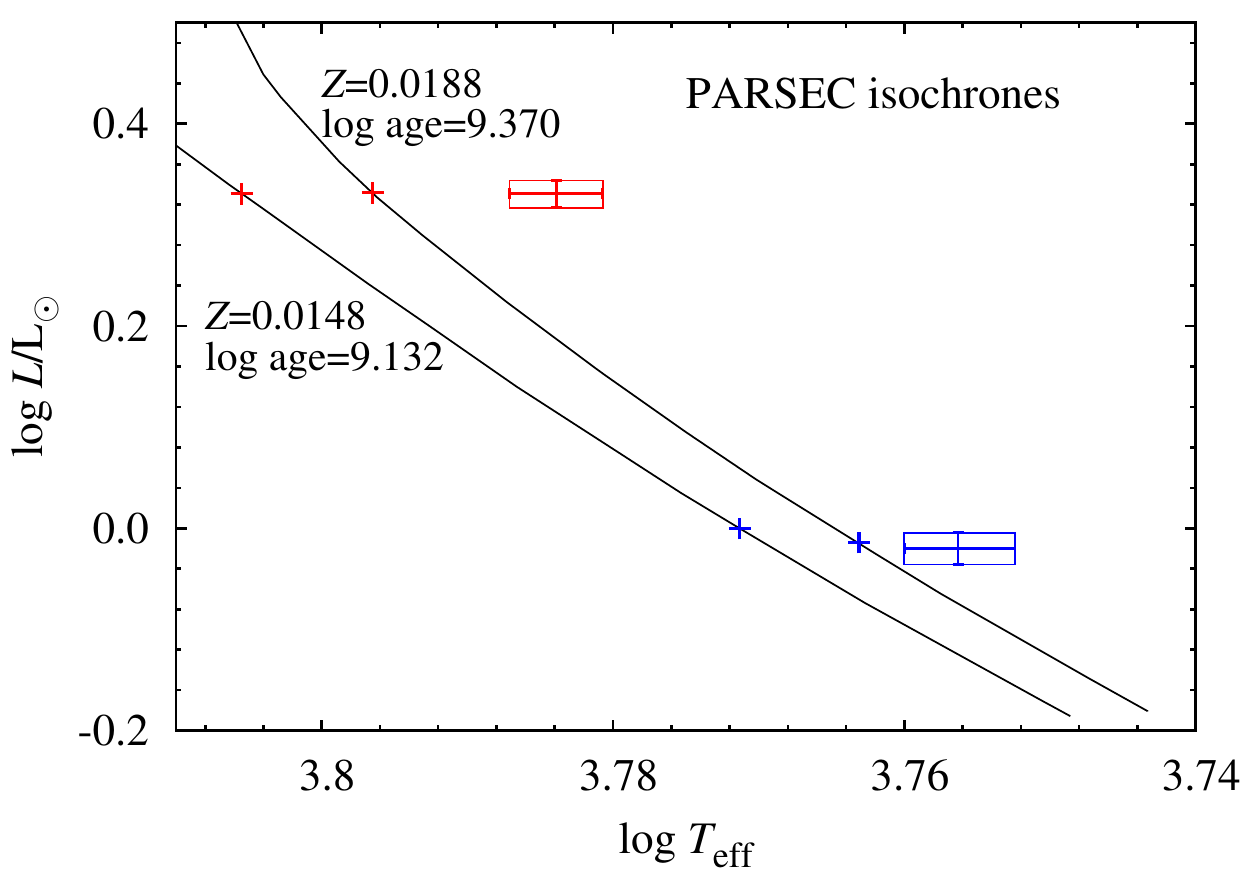}
\caption{Best-fitting PARSEC isochrones with $Z=0.0148$ and $Z=0.0188$. Upper error box denotes the position of the primary and the lower error box  of the secondary star. Small crosses denote positions of stars with masses equal to both components of LL Aqr.}
\label{fig:parsec}
\end{figure}

In Fig.~\ref{fig:parsec} we plot the best-fitting PARSEC isochrones for two metal abundances, $Z=0.0148$, which corresponds to the measured metal abundance of the components of the system, and for much higher metal content, $Z=0.0188$. Along each isochrone, we interpolate the point at which the masses are exactly equal to the masses of the components of LL~Aqr (as given in Table~\ref{par_fi}; see also below). Then, a particular isochrone (age) was selected to minimise the $\chi^2$ function in which we include the temperatures, radii, and luminosities of the two components. A severe disagreement is noticeable. Even for a very high metal abundance the isochrones are too hot.

The discrepancy is further confirmed with computation of stellar evolutionary tracks with the publicly available stellar evolution code MESA \citep[release 7184, e.g.][]{mesa3}. In the computations with MESA, we used OPAL opacities and solar heavy element distribution according to \cite[][hereafter AGSS09]{a09}. The atmospheric boundary condition was set through interpolation in the atmosphere tables, as described in \cite{mesa1}. Standard mixing-length theory was used \citep{mlt}. Convective overshooting was described following the standard approach with the extent of overshooting expressed as a fraction of the local pressure scale height, $\beta H_p$, measured above (or below) the border of the convective region determined with the Schwarzschild criterion. Only overshooting from the hydrogen burning core ($0.2H_p$) is included; as noted above the overshooting influences the primary only. In this section, the mixing-length parameter was set to $\alpha_{\rm MLT}=1.76$, which results from the calibration of the standard solar model. In the calibration, the overshooting from the convective envelope was neglected and element diffusion was included. Otherwise, exactly the same numerical set-up and microphysics data were used.

\begin{figure}
\centering
\includegraphics[width=9cm]{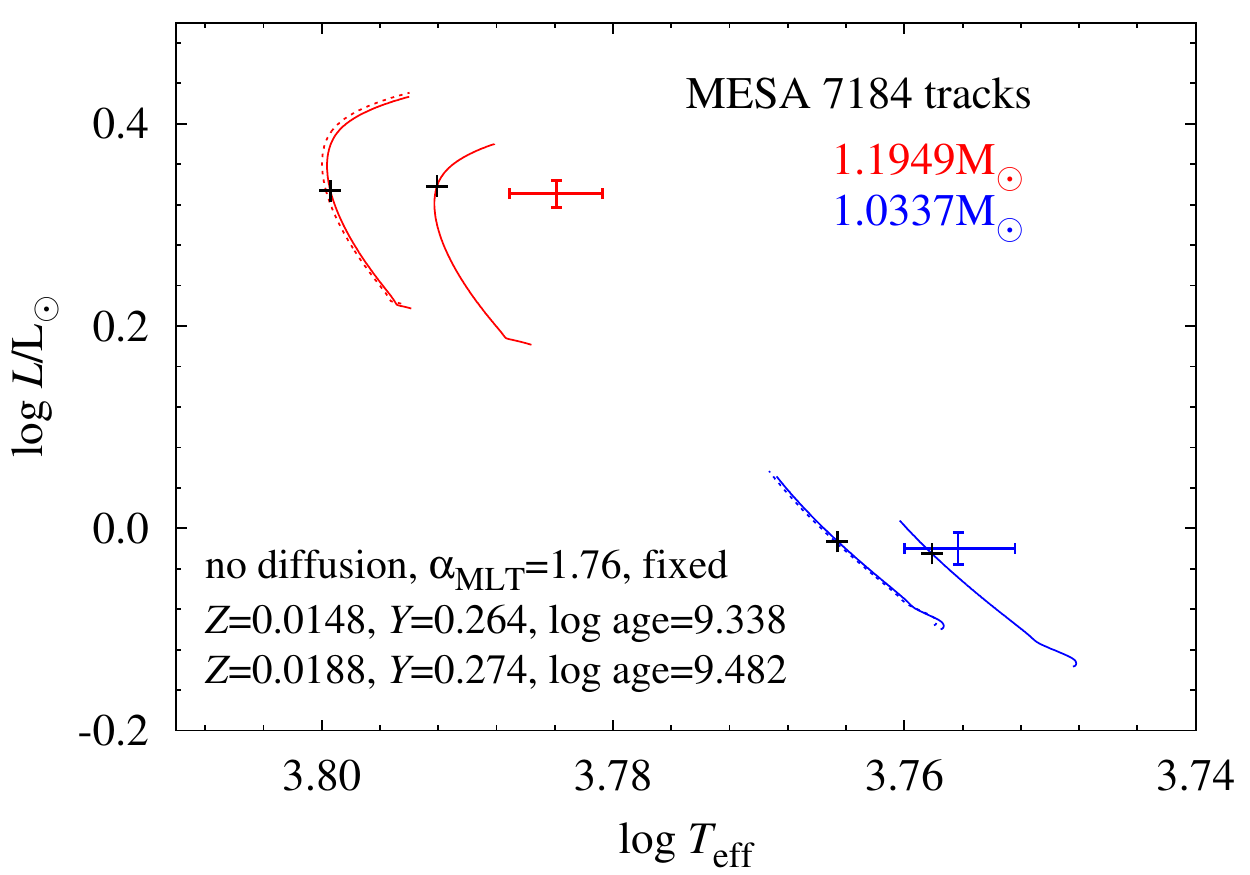}
\caption{Best-fitting MESA tracks with $Z=0.0148$ (left-side) and $Z=0.0188$ (right-side) and calculations with mixing-length parameter fixed to the solar-calibrated
value and element diffusion neglected. For $Z=0.0148$, the dotted tracks show the effect of increasing the masses of the components by $3\sigma$. Position of both components are indicated with errorbars. 
Small crosses denote position of the best fit in the three-dimensional space of parameters \{$T_{\rm eff}$, $L$, $R$\}.}
\label{fig:mesabasic}
\end{figure}

To model LL~Aqr, a small model grid with only two free parameters, the metal abundance, $Z\!\in\!(0.0138,\ldots,\,0.0188)$ (step $0.0005$), and the helium abundance, $Y\!\in\!(0.264,\ldots,\,0.276)$ (step $0.002$), was computed. Their surface values remain fixed as diffusion is neglected in the models considered in this section. The masses of the two components were fixed. The best models were selected by minimisation of $\chi^2$ (including temperatures, radii and luminosities; the latter two are not independent), which was calculated for each pair of tracks in the grid and at the same age of the components. The best matching models, assuming $Z=0.0148$ and $Z=0.0188,$ are plotted in Fig.~\ref{fig:mesabasic}. In general, the best fits, with very similar $\chi^2$ values, are obtained for $Z\geq 0.0178$ and various values of $Y$. The best fit for the highest metal abundance is plotted in Fig.~\ref{fig:mesabasic} to enable comparison with PARSEC isochrones.

The discrepancy between the models and observations is again apparent: the tracks are too hot. To get a better agreement (but still far from satisfactory) a much higher $Z$ is needed, than observed. The discrepancy is clearly larger for the primary. We note that at the same, $Z$, the agreement between models and observations, is much better for the MESA tracks. This is mostly due to the different helium abundance, which is tightly linked to $Z$ in the PARSEC isochrones ($Y\!=\!0.27472$ for $Z\!=\!0.0148$ and $Y\!=\!0.28210$ for $Z\!=\!0.0188$) and was a free parameter in the model grid computed with MESA. A much lower $Y$ in the latter case shifts the tracks towards cooler temperatures and makes the system older.

As noted above, in our calculations the masses of the components are fixed and equal to the values derived from the analysis of LL~Aqr observations. This is fully justified, taking into account the precise measurement of masses. In Fig.~\ref{fig:mesabasic} we show the effect of increasing the masses by $3\sigma$ for tracks with $Z=0.0148$ (dotted lines). The effect is barely visible; it is smaller than the effect of changing the metal content even by a small fraction of its formal measurement error.

\subsection{Advanced modelling of LL~Aqr}

Both PARSEC isochrones and standard MESA predictions indicate that the models are too hot compared to the observations. The disagreement is more severe for the primary. A significant increase in the metallicity of the system only reduces the discrepancy. In this section we explore the two effects that should improve the agreement with the observations without invoking too large a metallicity. The first effect is heavy element and helium diffusion. Its MESA implementation follows the seminal work by \cite{thoul}. Its inclusion allows us to set a higher initial metal abundance at the ZAMS, $Z_{\rm i}$, where the composition of both stars is homogeneous and is assumed to be the same. As stars evolve, the heavy elements sink owing to elemental diffusion and the photospheric $Z$ decreases. Diffusion is an efficient process in the Sun \citep[e.g.][]{bahcall}, hence, its inclusion in the modelling of LL~Aqr, composed of a solar twin and a slightly more massive primary, seems natural. Still, inclusion of microscopic diffusion is not a rule in stellar evolutionary calculations; it is sometimes neglected \citep[e.g.][]{stev} or only included in the calibration of the solar model \citep[e.g.][]{BaSTI}. The diffusion leads to the concentration of heavy elements towards the centre; however, the chemical composition is homogeneous in the outer convective envelope. Hence, the deeper the convective envelope is, the higher the envelope (and photospheric) metal abundance, $Z$. The overshooting from the convective envelope may thus affect the photospheric $Z$. The envelope is clearly deeper in the secondary, while it is very thin in the primary, as model calculations show. Therefore, it is expected that the photospheric $Z$ should be higher in the secondary. Although the effect is not observed, as the measurement errors, $Z_{\rm P}=0.0148(7)$
and $Z_{\rm S}=0.0149(11)$, clearly allow some difference between the metal abundance of the components.

Hydrodynamic simulations show that adopting a single value of the mixing-length parameter for models of different masses is not appropriate. Also, keeping $\alpha_{\rm MLT}$ fixed during evolution is not appropriate, as properties of the convection vary significantly across the HR diagram \citep[e.g.][]{ts11,ts14,mwa15}. While instantaneous adjustment of $\alpha_{\rm MLT}$ during the evolution of the model, using some prescription derived from hydrodynamic simulations, is beyond the scope of the present analysis, we can easily check the effects of adopting different $\alpha_{\rm MLT}$ values for the primary and for the secondary. An analysis of simulations presented in \cite{ts11}, \cite{ts14}, and \cite{mwa15} indicates that the primary's $\alpha_{\rm MLT}$ might be lower than secondary's $\alpha_{\rm MLT}$ by up to $\approx 0.1$. Since the secondary is not identical to the Sun, a slightly different value of $\alpha_{\rm MLT}$ than Sun's calibrated value might also be allowed.

Based on the above considerations we computed a large model survey for LL~Aqr with the following free parameters: the initial (ZAMS) metal abundance, $Z_{\rm i}\!\in\!(0.0148,\ldots,0.0183)$ (step $0.0005$, the same for both components), the initial ZAMS helium abundance, $Y_{\rm i}\!\in\!(0.258,\ldots,0.276)$ (step $0.002$; the same for both components), mixing-length parameter for the primary, $\alpha_{\rm P}\!\in\!(1.36,\ldots,1.76)$ (step $0.05$), mixing-length parameter for the secondary, $\alpha_{\rm S}\!\in\!(1.60,\ldots,1.76)$ (step $0.02$), and extent of the envelope overshooting, $\beta_{\rm e}\!\in\!(0,\ldots,0.8)$ (step $0.2$). Tests show that the extent of overshooting from the small convective core developed in the primary is not important (it is fixed to $\beta_{\rm H}=0.2H_p$). Masses of the two components are fixed to the values given in Table~\ref{par_fi}.

\begin{figure*}
\centering
\includegraphics[width=14cm]{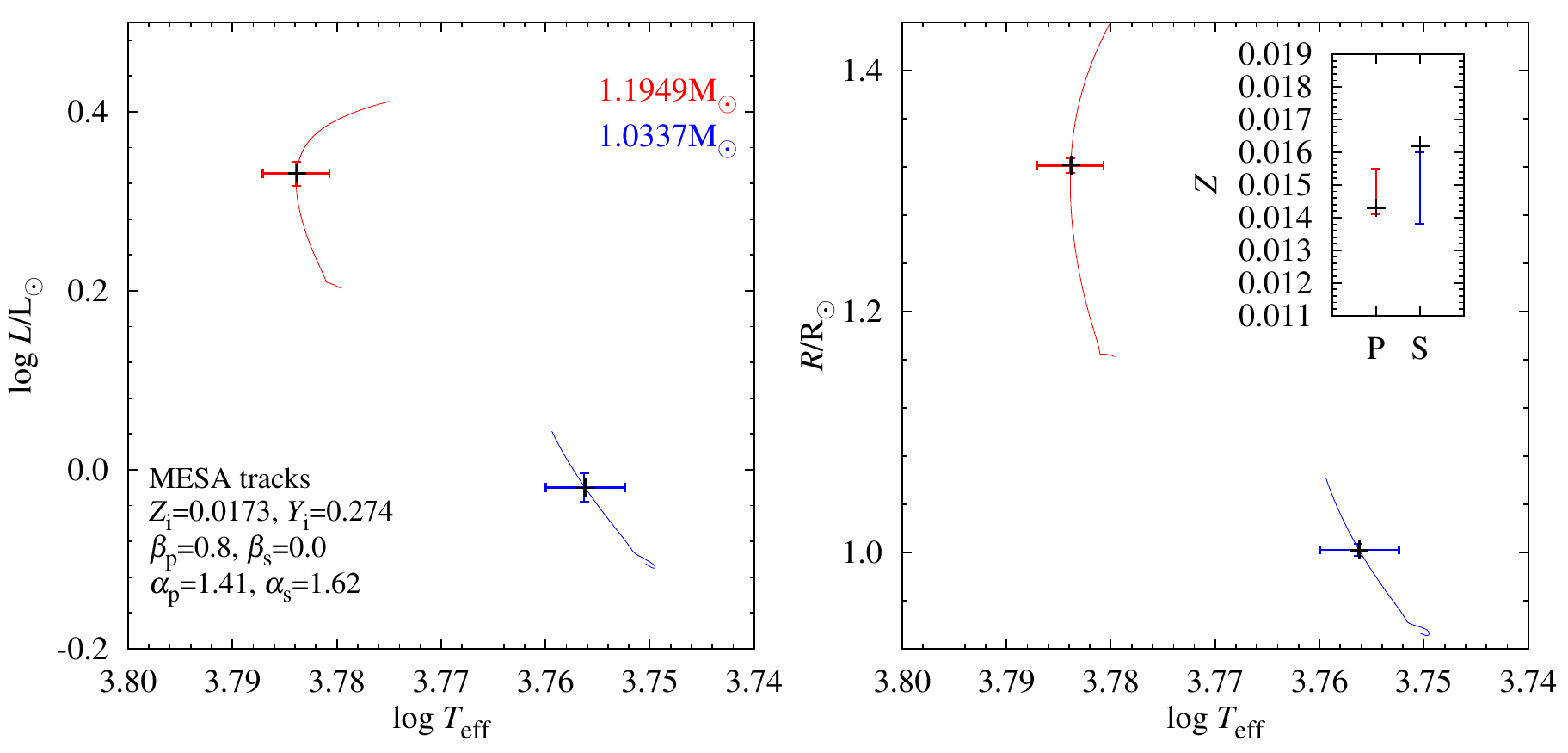} 
\caption{Best-fitting MESA tracks with element diffusion and allowing different values of mixing-length parameters for the models. No additional constraints are imposed on the models. Insert in the right panel shows comparison with observed metallicity abundance (P -- the primary, S -- the secondary). The meaning of the symbols is the same as in Fig.~\ref{fig:mesabasic}}
\label{fig:mesa_all}
\end{figure*}

We first discuss the best matching model in the considered grid without imposing any additional constraints. Because of the degeneracy between the model parameters (e.g.\ an increase of $Z_{\rm i}$, decrease of $Y_{\rm i}$, and decrease of $\alpha$ all have the same effect of making the tracks cooler) and because our parameter grid is relatively dense, there are many models with very similar $\chi^2$ value ($\chi^2$ now includes the metallicities of the components). In Fig.~\ref{fig:mesa_all} we plot the results for the best-fitting model ($\chi^2=2.3$). The initial, ZAMS, composition of the components is $Z_{\rm i}=0.0173$ and $Y_{\rm i}=0.274$. Radii, luminosities, and effective temperatures are matched nearly exactly at $\log{\rm age}=9.359$. The only significant discrepancy is observed for the current photospheric metal abundance of the secondary ($Z_{\rm S}=0.0162$), which is higher than observed by $\approx 1\sigma$ ;s ee insert in the right panel of Fig.~\ref{fig:mesa_all}. Although the fit is nearly perfect, the model parameters are difficult to accept. First, both mixing-length parameters are very low; the mixing-length parameter for the secondary, which is a solar twin, is $\alpha_{\rm S}=1.62$, which is far from the Sun-calibrated value ($1.76$). The difference between the mixing-length parameters of the components is also high, $\alpha_{\rm S}-\alpha_{\rm P}=0.21$, about twice as large as seems to be acceptable in the light of the hydrodynamic simulations quoted above. The envelope overshooting parameters are also hard to accept: no overshooting for the secondary, which has large convective envelope, and strong overshooting for the primary, which has a thin convective envelope. Such overshooting parameters reduce the difference in the current photospheric $Z$ of the components, but do not seem to be reasonable.

\begin{figure*}
\centering
\includegraphics[width=14cm]{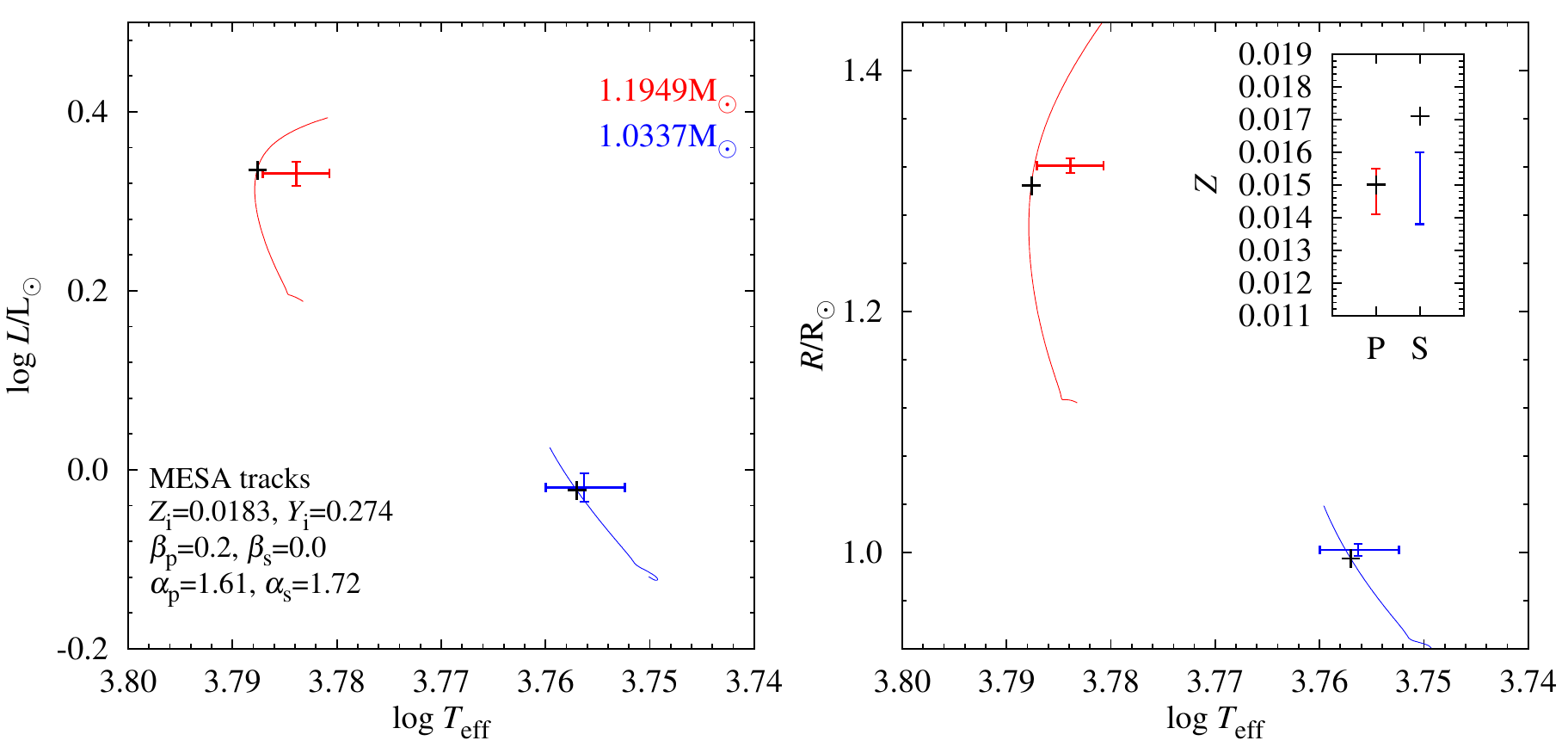}
\caption{Best-fitting MESA tracks with element diffusion and allowing different values of
mixing-length parameters for the models. Additional constraints were imposed on the models:
$\alpha_{\rm S}\geq 1.72$, $\alpha_{\rm S}-\alpha_{\rm P}\leq0.12$ and $|\beta_{\rm P}-\beta_{\rm S}|\leq 0.4$. The meaning of the insert is the same as in Fig.~\ref{fig:mesa_all}. The meaning of the symbols is the same as in Fig.~\ref{fig:mesabasic}}
\label{fig:mesa_con}
\end{figure*}

To check whether an acceptable fit can be obtained with more reasonable model parameters, we imposed the following constraints: (i) the mixing length for the secondary must be $\alpha_{\rm S}\geq 1.72$, (ii) the difference between the mixing length of the components must be $\alpha_{\rm S}-\alpha_{\rm P}<0.12$, and (iii) the difference in envelope overshooting of the components is not larger than 0.4. With these constraints, we find that the minimum $\chi^2$ values are around $30-35$ for several models with initial chemical compositions $Z_{\rm i}=0.0183$ or $Z_{\rm i}=0.0178$ and with $Y_{\rm i}$ in the $0.270-0.276$ range. Higher metallicities than in the previously discussed cases result from the constraint on the mixing-length parameters. Since these now cannot be arbitrarily low, tracks are shifted towards cooler temperatures (to match the properties of LL~Aqr) by the increase in metal abundance. The best model ($\chi^2=29.4$) is plotted in Fig.~\ref{fig:mesa_con}. The initial chemical composition at the ZAMS is $Z_{\rm i}=0.0183$, $Y_{\rm i}=0.274$ (for both components). The present compositions, at $\log{\rm age}=9.427$, are $Z_{\rm P}=0.0150$, $Y_{\rm P}=0.225,$ and $Z_{\rm S}=0.0171$, $Y_{\rm S}=0.255$. The metal abundance is $2.0\sigma$ larger than observed for the secondary. The mixing-length parameters are the lowest allowed by the imposed constraints. The extent of envelope overshooting for the secondary is $0.2H_p$, which is the lowest non-zero value considered in the grid. The overall fit for the secondary, except its metallicity, is very good. The fit is worse for the primary, but still reasonable; the tracks are about $1\sigma$ hotter and the match for metallicity is very good.

\subsection{Bayesian analysis with  \textsc{garstec} models}
\label{subsec:garstec}
 We also used the Bayesian method described in Kirkby-Kent et al. (2016, A\&A submitted) to compare the parameters of LL~Aqr to a large grid of stellar models calculated using the Garching
Stellar Evolution Code (\textsc{garstec}) \citep{2008ApSS.316...99W}. The methods used to calculate the grid are described by \cite{2015A&A...575A..36M} and \cite{2013MNRAS.429.3645S}.

{\textsc{garstec} uses the \cite{kww.book} mixing-length theory for convection, which for $\alpha_{ml} = 1.78$ produces the observed properties of the Sun assuming the composition given by \citet[][hereafter GS98]{1998SSRv...85..161G}.  Convective overshooting is treated as a diffusive process with overshooting parameter $f = 0.020$. Atomic diffusion of all atomic species is included by solving the multi-component flow equations of \cite{1969fecg.book.....B} using the method of \cite{thoul}. Macroscopic extra mixing below the convective envelope is  included following the parametrisation given in \cite{2012ApJ...755...15V}, which depends on the extension of the convective envelope. The initial helium abundance is calculated using
\begin{equation}
Y_{\rm i}  = Y_{\rm BBN} + Z_{\rm i}\frac{d Y}{d Z} + \Delta Y
\label{eq:HelAbunEq}
,\end{equation}
where \mbox{$Y_{\rm BBN} = 0.2485$} is the primordial helium abundance at the time of the Big Bang nucleosynthesis \citep{2010JCAP...04..029S}, \mbox{$ Z_{\rm i} $} is the initial metal content, and dY/dZ= 0.984 is calibrated using the values of the initial helium and metal content of the Sun  that provide the best fit to the properties of the present-day Sun (Y$_{\odot,i}= 0.26626$, Z$_{\odot,i} = 0.01826$, respectively). Further details of the models are described in Kirkby-Kent et al. (2016).

For this analysis we used model grids that cover three different mixing lengths, ($\alpha_{ml} =1.50, 1.78,$ and 2.04) for fixed initial helium abundance $\Delta Y = 0$, and five model grids with initial helium abundance $\Delta Y = 0, \pm0.01$, $\pm 0.02,$ and $\pm0.03$ with fixed mixing length $\alpha_{ml} = 1.78$. The mass range $0.6\,M_{\sun}$ to $2.0\,M_{\sun}$ is covered by the model grids in steps of $0.02\,M_{\sun}$, while the initial metallicity, [Fe/H]$_{\rm i}$, covers $-0.75$ to $-0.05$ in steps of $0.1$ dex and $-0.05$ to $+0.55$ in steps of $0.05$ dex.

The vector of model parameters used to predict the observed data is \mbox{$\vec{m} =(\mbox{$\tau_{\rm sys}$}, \mbox{$M_{1}$}, \mbox{$M_{2}$}, \mathrm{[Fe/H]}_{\mathrm{i}}$}, where $\mbox{$\tau_{\rm sys}$}$ is the age of the binary system, \mbox{$M_{1}$} and \mbox{$M_{2}$} are the stellar masses, and $\mathrm{[Fe/H]}_{\mathrm{i}}$ is the initial metal abundance.  The posterior probability distribution function of $\vec{m,}$ given the observed data $\vec{d}$, \mbox{$p(\vec{m}|\vec{d}) \propto {\cal L}(\vec{d}|\vec{m})p(\vec{m}),$} was determined using a MCMC method. The uncertainties on the mass, radius, and luminosity of both stars are correlated, which makes it awkward to calculate  the likelihood ${\cal L}(\vec{d}|\vec{m})$.
 Instead we use \mbox{\vec{d}=($T_{\rm eff,1}$, $\rho_{1}$,
$\rho_{2}$, $T_{\rm ratio}$, $M_{\rm sum}$, $q$, ${\rm [Fe/H]_{s}}$)}}, where $M_{\rm sum} = M_1 + M_2$, $q=M_2/M_1$,  $T_{\rm ratio} = T_{\rm eff,2}/T_{\rm eff1}$, and $\rho_{1,2}$ are the stellar densities. These parameters were chosen because they are closely related to a feature of the observatonal data and so are nearly independent, for example the stellar densities $\rho_1 = (0.5176 \pm0.0072)\rho_{\odot}$ and $\rho_2 = (1.026\pm 0.015)\rho_{\odot}$, are calculated directly from $R_1/a$ and $R_2/a$ via Kepler's third law. By assuming that these parameters are independent, we can calculate the likelihood using  \mbox{${\cal L}(\vec{d}|\vec{m}) = \exp(-\chi^2/2)$}, where
\begin{eqnarray} 
\chi^2   = & \left[\sum_{n=1,2}\frac{\left(\rho_{n} - \rho_{n,\mathrm{obs}}\right)^2}{\sigma_{\rho_{n}}^2} \right]
+\frac{\left(T_{\mathrm{eff, 1}} -T_{\mathrm{eff, 1,obs}}\right)^2}{\sigma_{T_{1}}^2}
+ \frac{\left(T_{\rm ratio} - T_{\rm ratio, \mathrm{obs}}\right)^2}{\sigma_{T_{\rm ratio}}^2} \\
\noalign{\smallskip}
& + \frac{\left(M_{\rm sum} - M_{\rm sum, \mathrm{obs}}\right)^2}{\sigma_{M_{\rm sum}}^2}
+ \frac{\left(q - q_{\mathrm{obs}}\right)^2}{\sigma_{q}^2}  + \frac{\left(\mathrm{[Fe/H]}_{\mathrm{s}} - \mathrm{[Fe/H]}_{\mathrm{s,obs}}\right)^2}{\sigma_{\mathrm{[Fe/H]_{\mathrm{s}}}}^2} \nonumber.
\end{eqnarray}

\noindent Observed quantities are denoted with `obs' subscript and their standard errors are given by the appropriately labelled values of $\sigma$.   We use a uniform prior for each model parameter over the available  grid range because these parameters are strongly constrained by the data so the choice of the prior probability distribution function, \mbox{$p(\vec{m})$}, has very little effect on the results. The initial point in the Markov chain is set using the best-fit age of the system for evolution tracks with masses fixed at the observed values and with [Fe/H]$_{\rm i}$ = [Fe/H]$_{\rm s}$. A ``burn-in'' chain of 50\,000 steps is used to improve the initial set of parameters and to calculate the covariance matrix of the model parameters. The eigenvectors and eigenvalues of this matrix are used to determine a set of uncorrelated, transformed parameters,  $\vec{q} = (q_1, q_2, q_3, q_4)$, where each of the transformed parameters has unit variance \citep{2004PhRvD..69j3501T}.  A final Markov chain of 500\,000 steps using these transformed parameters is then used to calculate \mbox{$p(\vec{m}|\vec{d})$}.

\begin{figure*}
\resizebox{\hsize}{!}{\includegraphics{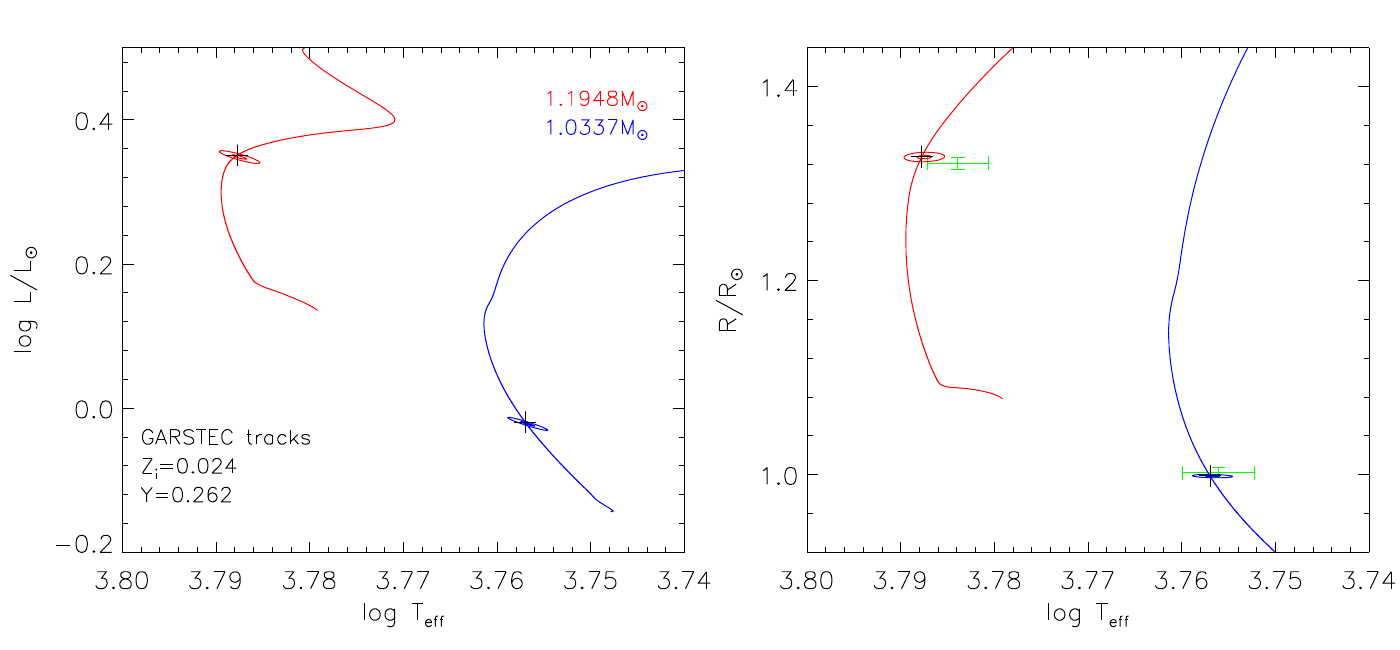}}
\caption{Best-fit  \textsc{garstec} stellar evolution tracks for LL Aqr. The posterior probability distributions for the parameters of both stars are shown using 1-sigma and 2-sigma error ellipses. The crosses show the same best-fit age for both tracks. In the right-hand panel the error bars show the adopted values for the effective temperature and radius of the stars.}
\label{GarstecFig}
\end{figure*}

The value of the effective temperature ratio has very little dependence on the T$_{\rm eff,1}$ so we use the value $T_{\rm ratio} =  0.9380 \pm0.0035$ for our analysis, where the estimate of the standard error comes directly from the error in $T_{\rm eff,2}$ in Table 5. For  [Fe/H]$_{s}$ we use the value for the primary star given in Table~6. The results of this analysis are given in Table~\ref{GarstecTable} and the best-fit evolution tracks for each star are compared to the posterior distributions for effective temparture ($T_{\rm eff}$), luminosity ($L$), and radius ($R$) in Fig.~\ref{GarstecFig}.

\begin{table*}
\caption{ {Age and parameters from the best-fitting model from a 500\,000-step Bayesian age fitting method for model grids with different mixing lengths and helium abundances. The mean and standard deviation of the resulting age distribution for each model grid are also shown.}}
\label{GarstecTable}
\centering
\begin{tabular}{r r c c c r r r r r r r r}
\hline\hline
\noalign{\smallskip}
${\alpha_{\rm ml}}$ & $\Delta Y$ & \multicolumn{1}{c}{$\tau_{\rm best}$} &  \multicolumn{1}{c}{$\tau_{\rm mean}$} & \multicolumn{1}{c}{$\sigma_{\tau_{\rm mean}}$} &  \multicolumn{1}{c}{$M_{1}$} &  \multicolumn{1}{c}{$M_{2}$} &  \multicolumn{1}{c}{${\rm [Fe/H]_{i}}$} &  \multicolumn{1}{c}{$T_{1}$} &  \multicolumn{1}{c}{$T_{2}$} &  \multicolumn{1}{c}{$\rho_{1}$} &  \multicolumn{1}{c}{$\rho_{2}$}         &  \multicolumn{1}{c}{$\chi^{2}$}  \\

 & & \multicolumn{1}{c}{(Gyr)} &  \multicolumn{1}{c}{(Gyr)} & \multicolumn{1}{c}{(Gyr)} &  \multicolumn{1}{c}{$(M_{\sun})$} &  \multicolumn{1}{c}{$(M_{\sun})$} &  &  \multicolumn{1}{c}{(K)} &  \multicolumn{1}{c}{(K)}  &  \multicolumn{1}{c}{($\rho_{\sun}$)} & \multicolumn{1}{c}{($\rho_{\sun}$)}   &  \\
\noalign{\smallskip}
\hline
\noalign{\smallskip}
1.50    & $0.00$        & $2.49$        & 2.49 & 0.11 & 1.1948 & 1.0336 & $0.10$ & 6098 & 5643 & 0.5252 & 0.9734 & 28.1     \\
1.78    & $0.00$        & $3.36$        & 3.37 & 0.13 & 1.1947 & 1.0336 & $0.14$ & 6157 & 5742 & 0.5129 & 1.0295 & 9.7  \\
2.04    & $0.00$        & $3.99$        & 3.99 & 0.14 & 1.1948 & 1.0337 & $0.16$ & 6187 & 5828 & 0.5087 & 1.0702 & 23.1     \\
\noalign{\smallskip}
\hline
\noalign{\smallskip}
1.78    & $-0.03$       & $4.37$        & 4.37 & 0.17 & 1.1948 & 1.0337 & $0.08$ & 6080 &5651 & 0.5070 & 1.0558 &12.2 \\
1.78    & $-0.02$       & $4.01$        & 4.02 & 0.16 & 1.1948 & 1.0337 & $0.09$ & 6111 &5685 & 0.5089 & 1.0467 & 9.1 \\
1.78    & $-0.01$       & $3.68$        & 3.68 & 0.14 & 1.1948 & 1.0337 & $0.12$ & 6131 & 5713 & 0.5106 & 1.0381 & 7.9  \\
1.78    & $0.00$        & $3.36$        & 3.37 & 0.13 & 1.1947 & 1.0336 & $0.14$ & 6157 & 5742 & 0.5129 & 1.0295 & 9.7  \\
1.78    & $+0.01$       & $3.06$        & 3.06 & 0.12 & 1.1948 & 1.0337 & $0.16$ & 6182 & 5772 & 0.5149 & 1.0220 & 14.5 \\
1.78    & $+0.02$       & $2.77$        & 2.77 & 0.11 & 1.1947 & 1.0336 & $0.18$ & 6211 & 5806 & 0.5165 & 1.0104 & 22.4 \\
1.78    & $+0.03$       & $2.49$        & 2.50 & 0.10 & 1.1948 & 1.0336 & $0.21$ & 6237 &5837 & 0.5190 & 1.0021 & 33.6 \\
\noalign{\smallskip}
\hline
\end{tabular}
\end{table*}

\section{Final remarks} \label{fin}

Careful re-examination of the radiative properties and absolute dimensions of both components of LL~Aqr confirmed our suspicions that the secondary is a very good solar twin candidate. Because of the favourable and simple geometric configuration, it is possible to derive both radii and masses very precisely. We were able to improve the radius measurements only slightly, but the mass determinations are greatly improved with respect to the results of previous works on the system. From all known solar twin candidates, LL~Aqr~B has its absolute dimensions known with by far the best precision.

Despite its apparent simplicity, theoretical modelling of LL~Aqr with PARSEC and MESA codes turned out to be very challenging. Standard isochrones and stellar evolutionary tracks (mixing-length parameters fixed to solar calibrated value) cannot reproduce the position of LL~Aqr in the HR diagram; the models are too hot. The discrepancy is more severe for the more massive primary component. Two effects clearly improve the agreement: inclusion of element diffusion in the stellar evolutionary tracks and allowing for different mixing-length parameters for the two components. Inclusion of both effects is fully justified, the latter based on recent hydrodynamic simulations. The nearly perfect match (Fig.~\ref{fig:mesa_all}) is, however, obtained for model parameters that do not seem reasonable. The mixing-length parameter for solar twin (secondary) is much lower than for the Sun; also, the difference between the mixing lengths of the two components is difficult to reconcile with the results of hydrodynamic simulations because it is much too large. When additional constraints are put on the models, to avoid the aforementioned difficulties the best matching models 
are still reasonable (Fig.~\ref{fig:mesa_con}). The most significant discrepancies concern the metallicity of the secondary and effective temperature of the primary; both are too low as compared with the models (effective temperature by $1\sigma$ only). 

These calculations clearly demonstrate the power of main-sequence eclipsing binaries with precisely determined parameters in testing stellar evolution theory. With a larger sample of such systems, parameters such as element diffusion and the efficiency of convective energy transfer, as a function of the location of a star in the HR diagram, could be studied. Based on modelling only LL~Aqr, we can conclude that a better treatment of convection than standard mixing-length theory, is necessary, which is also obvious in light of the hydrodynamical simulations. The treatment of diffusion is also still uncertain; see for example\ the discussions in \cite{parsec} and \cite{PISA}. When confronted with precise observables, stellar evolution theory is clearly deficient. The best example is the modelling of the Sun. The depth of the convective envelope and surface helium abundance in the standard solar model disagree with the very precise asteroseismic measurements \citep{basu}. Whatever the cause, it will also affect LL~Aqr and similar systems. There is a growing amount of evidence that an increase in the opacity coefficient is needed in the so-called metal opacity bump (Z-bump), which is located at a temperature of $\log{T} \approx 5.3$ in a stellar atmosphere and caused by a large number of absorption lines produced by fine-structure transitions in the ions of the iron group.
Such the increase is needed, not only to improve the modelling of the Sun \citep[e.g.][]{serenelli}, but also to improve the modelling of other pulsating stars, for example\ the B-type pulsators \citep[e.g.][]{salmon,walczak}. It will also help to construct better models for LL~Aqr. Now, to get the better agreement, a higher metallicity than observed is needed. The increase in $Z$ mimics the increase of the $Z$-bump opacities. Hence, the increase in the $Z$-bump opacities would also improve the agreement between the models and observations without invoking metallicities that are  too large.

Comparison of results from modelling LL Aqr using {\sc garstec} and MESA stellar evolution codes show both similarities and differences. The two codes predict hotter and more metal rich components by about 1$\sigma$ and 1.5$\sigma$, respectively. However, the \textsc{garstec} code predicts a significantly older age of the system (see Tables~\ref{par_fi}), higher initial metallicity, and suggests some slight helium underabundance. 
It can be explained as follows. As {\sc garstec} models use GS98 solar mixture and MESA models use AGSS09, this leads to {\sc garstec} models being
hotter by about 200 K if $\alpha_{\rm MLT}$, $Y$ and $Z$ are fixed (A. Serenelli, private communication). 
In order to make {\sc garstec} tracks cooler one needs to both an increase in $Z$ and
reduction in $Y$. This has the consequence of making tracks less
luminous, so the observed luminosities are achieved at
later evolutionary stage, i.e. larger ages. Anyway it would be possible to improve this fit by exploring the influence of other free parameters, such as the extent of the extra macroscopic mixing below the convective envelope, that are included in these models.

The future work on the system should consist in a very detailed differential analysis of a spectrum of the secondary in respect of the solar spectrum.  A high quality decomposed spectrum is needed with S/N of at least 200 to make this analysis feasible. That would allow us to redetermine  atmospheric parameters such as temperature and metal abundance more accurately, and answer the question of whether tensions with evolution models predictions are caused by some systematics in our atmospheric parameters determination (especially temperature) or are caused by some deficiencies of the evolutionary codes. The system is also well suited for determination of its astrometric orbit with interferometry \citep[e.g.][]{gall16}: maximum angular separation between components is 1.78 mas and the $H$-band flux ratio should be 0.53. Resulting geometric distance will allow for deriving very precise angular diameters of the components and for improving surface brightness colour calibrations.   

\begin{acknowledgements}

Support from the Polish National Science Center grants MAESTRO DEC-2012/06/A/ST9/00269 and OPUS DEC-2013/09/B/ST9/01551 is acknowledged. We [D.G., W.G., G.P., M.G.] also gratefully acknowledge financial support for this work from the BASAL Centro de Astrofisica y Tecnologias Afines (CATA) PFB-06/2007, and from the Millenium Institute of Astrophysics (MAS) of the Iniciativa Cientifica Milenio del Ministerio de Economia, Fomento y Turismo de Chile, project IC120009. The work of KP has been supported by the Croatian Science Foundation under grant 2014-09-8656. Fruitful discussions with Wojtek Dziembowski and Pawel Moskalik are acknowledged. The Leverhulme Trust is acknowledged for supporting the work of JS. PFLM is supported by funding from the UK's Science and Technology Facilities Council. We are much indebted to Aldo Serenelli for discussion of {\sc garstec} code.

We used SIMBAD/Vizier database in our research. We used also the {\it uncertainties} python package.

\end{acknowledgements}

\label{lastpage}

\end{document}